\documentclass[letterpaper]{article}

\PassOptionsToPackage{numbers,sort&compress,square}{natbib}
\usepackage[preprint]{neurips_2025}

\usepackage[utf8]{inputenc} 
\usepackage[T1]{fontenc}    
\usepackage[hidelinks]{hyperref}       
\usepackage{url}            
\usepackage{booktabs}       
\usepackage{amsfonts}       
\usepackage{nicefrac}       
\usepackage{microtype}      
\usepackage{xcolor}         
\usepackage{orcidlink}
\usepackage{pdfpages}
\usepackage{tabularx,booktabs,ragged2e,makecell,adjustbox,caption,placeins}
\newcolumntype{Z}{>{\RaggedRight\arraybackslash}X} 
\newcolumntype{Y}{>{\centering\arraybackslash}X}   

\title{Chat to Chip: Large Language Model Based Design of Arbitrarily Shaped Metasurfaces}

\author{%
	Huanshu Zhang \orcidlink{0009-0009-8332-7298}\\
	Department of Electrical Engineering\\
	The Pennsylvania State University\\
	University Park, PA 16802 \\
	\texttt{hpz5226@psu.edu} \\
	\And
	Lei Kang \orcidlink{0000-0001-7718-7756}\\
	Department of Electrical Engineering\\
	The Pennsylvania State University\\
	University Park, PA 16802 \\
	\texttt{lzk12@psu.edu} \\
	\And
	Sawyer D. Campbell \orcidlink{0000-0002-3973-2730}\\
	Department of Electrical Engineering\\
	The Pennsylvania State University\\
	University Park, PA 16802 \\
	\texttt{sdc22@psu.edu} \\
	\And
	Douglas H. Werner \orcidlink{0000-0001-5629-6478}\\
	Department of Electrical Engineering\\
	The Pennsylvania State University\\
	University Park, PA 16802 \\
	\texttt{dhw@psu.edu} \\
}

\makeatletter
\def\bstctlcite#1{\@bsphack
	\@for\@citeb:=#1\do{%
		\edef\@citeb{\expandafter\@firstofone\@citeb}%
		\if@filesw\immediate\write\@auxout{\string\citation{\@citeb}}\fi}%
	\@esphack}
\makeatother

\begin{document}
	\bstctlcite{BSTcontrol}
	
	\maketitle
	\begingroup
	\renewcommand\thefootnote{}\footnotetext{Accepted manuscript at \emph{Nanophotonics}. DOI: \url{https://doi.org/10.1515/nanoph-2025-0343}. The Version of Record may differ slightly.}
	\addtocounter{footnote}{-1}
	\endgroup

	\begin{abstract}
		Traditional metasurface design is limited by the computational cost of full-wave simulations, preventing thorough exploration of complex configurations. Data-driven approaches have emerged as a solution to this bottleneck, replacing costly simulations with rapid neural network evaluations and enabling near-instant design for meta-atoms. Despite advances, implementing a new optical function still requires building and training a task-specific network, along with exhaustive searches for suitable architectures and hyperparameters. Pre-trained large language models (LLMs), by contrast, sidestep this laborious process with a simple fine-tuning technique. However, applying LLMs to the design of nanophotonic devices, particularly for arbitrarily shaped metasurfaces, is still in its early stages; as such tasks often require graphical networks. Here, we show that an LLM, fed with descriptive inputs of arbitrarily shaped metasurface geometries, can learn the physical relationships needed for spectral prediction and inverse design. We further benchmarked a range of open-weight LLMs and identified relationships between accuracy and model size at the billion-parameter level. We demonstrated that 1-D token-wise LLMs provide a practical tool to designing 2-D arbitrarily shaped metasurfaces. Linking natural-language interaction to electromagnetic modelling, this “chat-to-chip” workflow represents a step toward more user-friendly data-driven nanophotonics. 
	\end{abstract}
	\paragraph{Keywords} Metasurfaces; Large Language Model; Deep Learning.

	\section{Introduction}
	Metasurfaces, which are defined as planar arrays of subwavelength scatterers that modulate the amplitude, phase, and polarization of light locally, have quickly become pivotal to nanophotonic devices \citep{cui_roadmap_2024}, enabling applications from high-numerical-aperture meta-lenses \citep{khorasaninejad_metalenses_2017} and holographic imagers \citep{huang_metasurface_2018} to augmented-reality displays \citep{ding_waveguide-based_2023}.  Despite this progress, metasurface design remains constrained by the need for brute-force full-wave electromagnetic solvers such as the finite-difference time-domain (FDTD) \citep{hao_fdtd_2009} and finite-element methods (FEM) \citep{silvester_finite_1996}. A single design iteration must traverse a high-dimensional parameter space, carry out numerous simulations, and finely adjust geometric features to satisfy spectral and/or wave-front requirements \citep{elsawy_numerical_2020}. For practical devices targeting large apertures and multiple functionalities, the corresponding computational load may take days or weeks, even when executed on large clusters or supercomputers \citep{li_advances_2020}. The resulting limitation discourages the exploration of unconventional materials, multilayer stacks, and fully aperiodic layouts. To keep pace with the burgeoning applications for metasurfaces, new design paradigms that bypass repeated heavy-duty simulations are urgently required.
	
	Recent breakthroughs in data-driven modelling offer a promising alternative route \citep{campbell_advances_2023}. Once trained on curated pairs of optical or electromagnetic responses with corresponding metasurface geometry, deep neural networks (DNNs) can predict the optical response of previously unseen geometries within milliseconds, marking a structure evaluation orders of magnitude faster than that based on full-wave solvers \citep{khaireh-walieh_newcomers_2023}, \citep{lee_datadriven_2024}. Recent works have shown the potential of DNN-based approaches for metasurface design \citep{an_deep_2022}, \citep{zhang_diffusion_2023}, \citep{campbell_explosion_2021}. For instance, Malkiel et al. employed a DNN for H-shaped plasmonic nanostructure design \citep{malkiel_plasmonic_2018}. An et al. developed a DNN to predict wideband amplitude and phase responses of quasi-freeform dielectric metasurfaces \citep{an_deep_2020}. Chen et al. introduced a transformer-based model for both forward and inverse design of broadband solar metamaterial absorbers \citep{chen_broadband_2023}. Moreover, Zhang et al. proposed a fixed-attention mechanism for the design of high-degree-of-freedom metamaterials \citep{zhang_fixed-attention_2025}.
	
	Although DNN-based models have demonstrated impressive accuracy and speed, integrating them into a metasurface design pipeline is still far from a turnkey off-the-shelf procedure \citep{zhang_data_2025}. Each new optical function typically requires a new training set, a custom network topology, and exhaustive hyper-parameter selections. This typically includes choosing the number of layers and neurons in each layer, which is an iterative, code-heavy process driven largely by heuristic intuition rather than first-principles guidance \citep{lakhmiri_hypernomad_2019}. To this end, Large Language Models (LLMs) present a qualitatively different proposition. LLMs are transformer-based neural networks that encapsulate billions of parameters in a single, frozen architecture pre-trained on vast amounts of natural-language text and code \citep{chang_survey_2024}. In this stage, the model is taught the simple objective of predicting the next word in a sequence; yet, by doing so at web-scale, it internalizes syntax, semantics, and a surprising amount of factual and mathematical structure \citep{naveed_comprehensive_2024}. Since the core model is fixed, researchers can simply train the LLM on task-specific datasets instead of re-designing and re-training a new network for every new task, thereby eliminating the laborious network-sizing and hand-tuning that DNNs demand. These characteristics make LLMs ideal candidates for enabling efficient design of metasurfaces with complex structures and layouts that possess various targeted functionalities.
	
	When domain precision is required, the same model can be “further trained” in an efficient manner (i.e., fine-tuned) on a relatively small, task-specific dataset, such as predicting the transmission spectrum of a metasurface \citep{dinh_lift_2022}, \citep{hu_lora_2021}. In practice, this dataset pairs sequence-based descriptions of each unit cell (geometric parameters, material indices, lattice spacing, and so on) with descriptions of its simulated optical response (spectral magnitude, phase, near-field maps, etc.). This pairing mirrors the input-output structure of conventional DNNs but represents both geometry and responses in a language-like format amenable to LLMs. Because LLMs accept byte streams, neither architectural redesign nor feature engineering is necessary: the model simply learns the mapping of geometries to responses in passes. After fine-tuning on a new dataset, which takes slightly longer than the time taken to train one custom DNN, the LLMs can predict spectra within seconds, providing near-real-time feedback during design loops while removing the code-heavy scaffolding and exhaustive hyper-parameter sweeps that traditional DNN-based methods demand. Thus, LLMs promise a “chat-to-chip” route for modelling metasurfaces. For example, in their pioneering study Kim et al. fine-tuned Llama \citep{grattafiori_llama_2024} for both forward prediction and inverse design of all-dielectric metasurfaces \citep{kim_nanophotonic_2025}, lowering the entry barrier for researchers who lack machine-learning background. Lu et al. fine-tuned ChatGPT 3.5 on various details of prompts and temperatures for the design of metamaterials \citep{lu_learning_2025-1}, Liu et al. used LLMs for design recommendation of phosphorescent materials \citep{liu_design_2025}, and, by optimizing and stitching wavelength-scale superpixels, Lupoiu et al. introduced a multi-agent LLM framework paired with a surrogate Maxwell solver that autonomously designs metasurfaces in near-real time \citep{lupoiu_multi-agentic_2025}. However, scaling these approaches from parameterized meta-atoms to arbitrarily shaped metasurfaces is of great importance to numerous applications but remains largely unexplored \citep{whiting_meta-atom_2020}, \citep{jenkins_establishing_2021}. Token-wise attention is intrinsically one-dimensional, whereas free-form surfaces require rich spatial reasoning. Emerging hybrids that couple LLM backbones with graph or vision transformers, or that embed topology as structured tokens, may be a possible a solution \citep{tang_graphgpt_2024}. However, a generally applicable framework for LLM-accelerated design of complex metasurfaces has yet to be reported, as accuracy of vision language models (VLMs) still lags behind LLM-level reliability while their implementation is cost-intensive and fragile \citep{yin_survey_2024}.
	
	Here, we present a workflow using LLMs to accelerate both forward and inverse design of arbitrarily shaped metasurfaces. We note that in our study an “arbitrarily shaped” meta-atom refers to a planar structure with a non-canonical or free-form shape, rather than a fully unparameterized one. Although limited to one-dimensional token streams, our results show that sequence-based LLMs are capable of capturing the physics required to predict optical responses for arbitrarily shaped metasurfaces. Also, for the inverse design section, our workflow addresses the designs of high-degree-of-freedom, randomly shaped 2D unit cells, which cannot be solved by existing image-generation or multimodal LLM approaches. This method eliminates DNN engineering and therefore further lowers the barrier for researchers with limited expertise in machine learning. Finally, cross‑model benchmarks that exploit state‑of‑the‑art LLMs in this design task are provided, establishing reference baselines to guide future work on LLM-accelerated photonic design.

	\section{Methods}
	Figure 1a outlines the workflow we use to generate arbitrarily shaped meta-atoms, a successfully verified parameterization approach adopted from \citep{whiting_meta-atom_2020}. First, a $4\times4$ control-point grid was randomly generated, where each element ranges within $[0, 1]$. The grid was replicated by a four-fold rotational symmetry, yielding a $7\times7$ lattice of control values. Interpolation converts these discrete values into a $256\times256$ surface. Binarization at a fixed threshold ($t = 0.5$) converts this surface into a preliminary foreground-background mask. To ensure the pattern is fabrication-friendly, the mask undergoes iterative morphological opening and closing until no further topological changes occur \citep{jenkins_establishing_2021}. This regularisation step eliminates isolated islands, fills holes, and enforces a minimum feature width and gap size compatible with standard fabrication processes. The final design is a $1000\,\mathrm{nm}\times1000\,\mathrm{nm}$ square unit cell comprising an arbitrarily shaped silicon pattern (refractive index $= 3.5$) generated using this approach sitting on a glass substrate (refractive index $= 1.5$).
	
	We generated a dataset of 45{,}790 metasurface designs with randomly generated control-point grids illuminated by a left-handed circularly polarized (LCP) normally incident wave. These designs were simulated using the commercial software package Lumerical FDTD on a server with two Intel(R) Xeon(R) Gold 6258R CPUs and 1.5TB memory. After every simulation, the transmission spectrum was recorded at 31 uniformly spaced wavelengths from $1050\,\mathrm{nm}$ to $1600\,\mathrm{nm}$. The completed dataset was randomly partitioned into training and test sets with a 4:1 ratio, resulting in 36{,}632 training samples and 9{,}158 test samples.
	
	To prepare the geometrical-optical pairs for the forward-prediction task using LLMs, each $4\times4$ control-point grid is converted into a natural-language prompt and its 31-element transmission vector into the corresponding completion. A typical prompt would be: “We have a 4-by-4 grid: [[g11, …, g14], …, [g41, …, g44]], what is the transmission spectrum of the metasurface generated using this grid?” while the target completion would then be: “The transmission values sampled at 31 evenly spaced points between 1050 nm and 1600 nm for the metasurface generated using this grid are [t1, …, t31]”. All numerical values are rounded to three decimal places, a choice that is not accuracy-limiting at the error scales considered, balancing GPU memory usage with predictive accuracy, and aligns with prior works \citep{kim_nanophotonic_2025,lu_learning_2025-1}. All prompts and expected outputs are tokenised with the same byte-pair encoder as the base model to ensure vocabulary consistency. Fine-tuning proceeds by feeding these prompt-completion pairs to the LLM and minimising the loss between the predicted tokens and the ground-truth completion, as illustrated in Figure 1b. This formulation re-casts spectrum prediction as language-sequence completion, allowing us to exploit the LLMs’ autoregressive training objective without architectural modifications.
	
	The LLM implemented in both the forward and inverse design process is Meta-Llama-3.1-8B-Instruct \citep{grattafiori_llama_2024} quantised to 4-bit weights by Unsloth. This LLM is identical to that employed by Kim et al.\ \citep{kim_nanophotonic_2025}, eliminating the need for neural network engineering. Parameter-efficient adaptation is realised with Low-Rank Adaptation (LoRA \citep{hu_lora_2021}), which injects low-rank adapters into all projection layers. The entire workflow is built using open-source libraries including Pytorch, HuggingFace, and Unsloth. All 7–9B parameter LLMs are trained on a single NVIDIA RTX 2080 Ti GPU, while the larger and smaller LLMs used in the benchmarking process are trained on one NVIDIA L40S GPU. The rank $r$ and the scaling factor $\alpha$ were both set to 32. Fine-tuning proceeds for 8 epochs with an effective batch size of 192, using the AdamW optimiser with an initial learning rate of $4.0\times10^{-4}$ followed by linear decay and the standard cross-entropy objective for next-token prediction. This setup resulted in approximately 10GB of GPU memory usage for 7–9B models, a requirement that is met by most contemporary commercially available consumer graphics cards.
	
	\begin{figure}
		\centering
		\includegraphics[width=\linewidth]{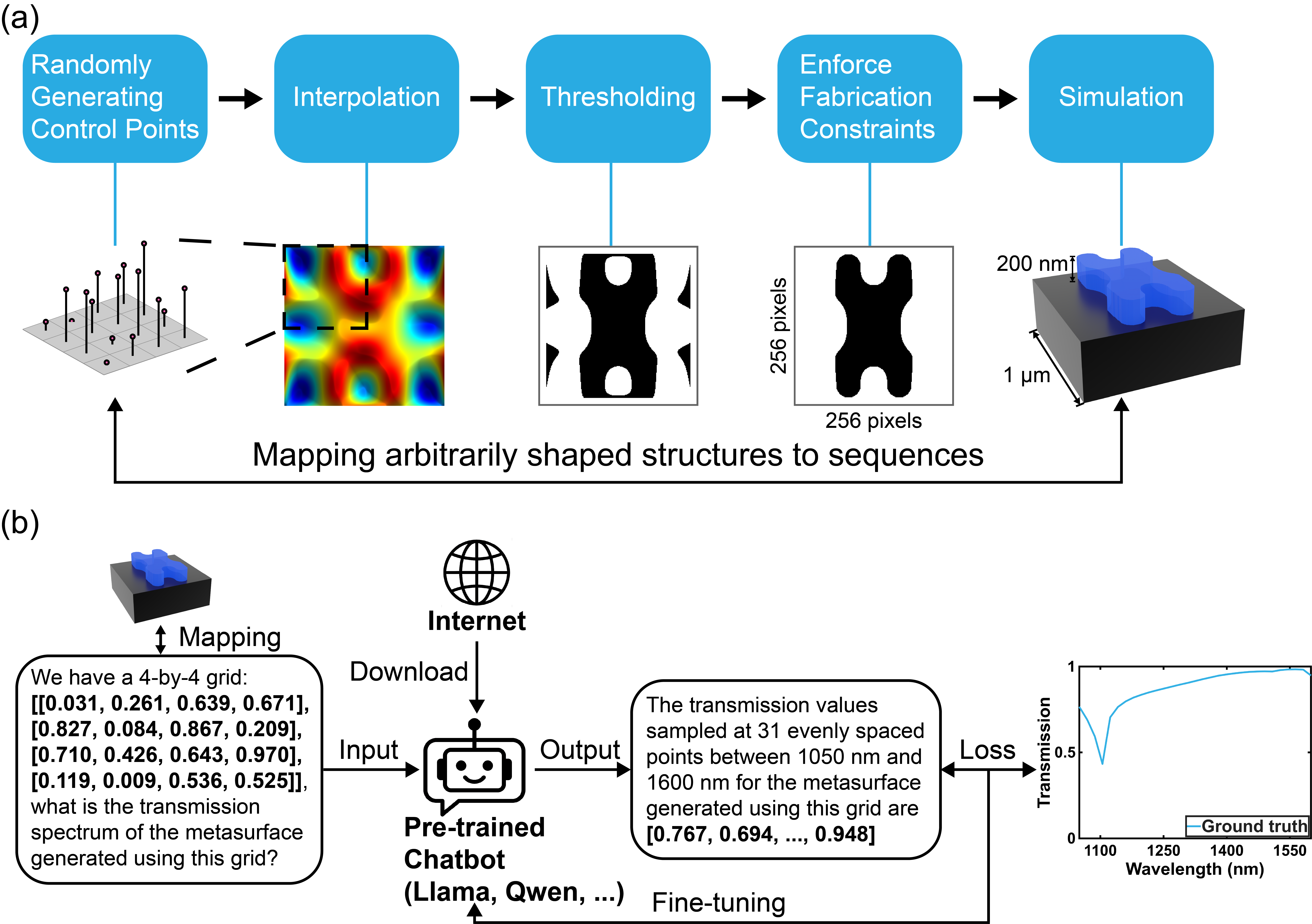} 
		\caption{Mapping arbitrarily shaped metasurface geometries to language sequences and training an LLM for rapid optical prediction. (a) A $4 \times 4$ matrix of a randomly sampled grid of control-points is replicated by four-fold rotational symmetry, interpolated into a $256 \times 256$ scalar field, binarized at a fixed threshold of 0.5, and regularised by iterative morphological opening/closing that removes isolated features smaller than 8,192 pixels and seals internal voids. The resulting binary mask is then extruded into a 200 nm-thick silicon layer on a 1 µm-pitch glass substrate and analysed with FDTD, establishing paired grid-spectrum data. (b) Fine-tuning and inference process for forward prediction. Each grid-spectrum pair is rewritten as a natural-language prompt that encodes the control-point grid and a target output that lists the 31 transmission values between 1,050 nm and 1,600 nm. Moreover, parameter-efficient fine-tuning (LoRA) of a pre-trained LLM minimises cross-entropy between predicted and ground-truth tokens, so that at inference the model returns an accurate spectrum within seconds from a single grid prompt, eliminating the need for labour-intensive network design.}
		
	\end{figure}

	\section{Results and discussion}
	\subsection{Forward design}
	To demonstrate the prediction accuracy, Figure 2 compares the spectra predicted by our fine-tuned Llama-3.1-8B with FDTD simulation results for four representative meta-atoms. The orange dashed curves (Llama) track the blue solid curves (FDTD) almost perfectly across the 1000-1600 nm band, faithfully reproducing both plateaus and sharp resonances. Querying the model is straightforward: copy and paste the $4\times 4$ control-point matrix into a prompt, as discussed previously, and then reading out the 31-point spectrum returned by the LLM. On a single RTX 2080 Ti GPU, this prediction takes approximately 2 seconds, about 60 times faster than the corresponding full-wave simulation on our CPU cluster. Across the entire 9158 sample test set the mean squared error (MSE) is $3.4\times 10^{-3}$ when trained for 8 epochs, matching specialised DNNs reported in the literature \citep{jenkins_establishing_2021,dong_advanced_2025}. In other words, a lightly fine-tuned (i.e., only need to tune the epochs number), off-the-shelf LLM delivers turnkey, high-fidelity forward modelling without any bespoke network design or hyper-parameter sweeps, demonstrating a practical “no code” path to rapid metasurface prototyping.
	
	Furthermore, to quantify the influence of fine-tuning epochs, we tested how different training epochs impact the prediction accuracy. Figure 3a confirms that fine-tuning length is a minor knob for this prediction task. Specifically, when the Llama-3.1-8B model is fine-tuned for between 5 and 20 epochs, a typical range for fine-tuning LLM, its test-set MSE meanders within $3.4 - 4.7\times 10^{-3}$, comfortably below the $5\times 10^{-3}$ marked by the red dashed line. In contrast, a hand-built eight-layer fully connected network swings from a best-case $2.0\times 10^{-3}$ to $1.78\times 10^{-2}$ after enlarging each layer from 512 to 895 neurons, marked by the two dark blue dash lines. Custom DNNs can still edge out the LLMs by a small margin, but only at the cost of exhaustive architecture searches. Hence, a pragmatic workflow is to deploy LLMs for rapid evaluation and reserve heavyweight solvers or customized networks for final, high-precision refinement stages. We note that the proposed workflow is excitation-agnostic: adapting to other illumination conditions such as different polarization or incidence angle simply requires changing the simulation setup and regenerating the corresponding dataset with no architectural changes required \citep{jiang_global_2019}.
	
	However, not all instruction-tuned LLMs are well-suited for numerical regression. When we fine-tuned Qwen-3B, a reasoning model, it repeatedly abandoned the target format and instead produced a verbose diagnostic such as: “\textit{To determine the transmission spectrum of the metasurface generated using the given 4-by-4 grid, we need to analyze the optical properties of the grid. Each value in the grid represents the transmission coefficient for a specific wavelength or frequency. However, the provided grid does not include information about the wavelength range or the specific frequencies corresponding to each value. Additionally, the transmission spectrum typically requires information about the incident light, the material properties, and the geometry of the metasurface. Without this information, it is not possible to accurately compute the transmission spectrum. Please provide additional details about the wavelength range, incident light parameters, and material properties to proceed with the calculation.}” Other similar reasoning LLMs, such as Phi-4-Reasoning, Llama-4, gpt-oss, and Gemma-3, exhibit similar behaviours during our fine-tuning. Such chain-of-thought digressions reveal that strong conversational priors can overshadow the supervised objective, prompting the model to seek more information from the user rather than produce the requested 31 transmission values. Therefore, reasoning-centric LLMs may demand additional engineering before they serve reliably as high-throughput, numeric predictors in scientific design loops.
	
	\begin{figure}
		\centering
		\includegraphics[width=\linewidth]{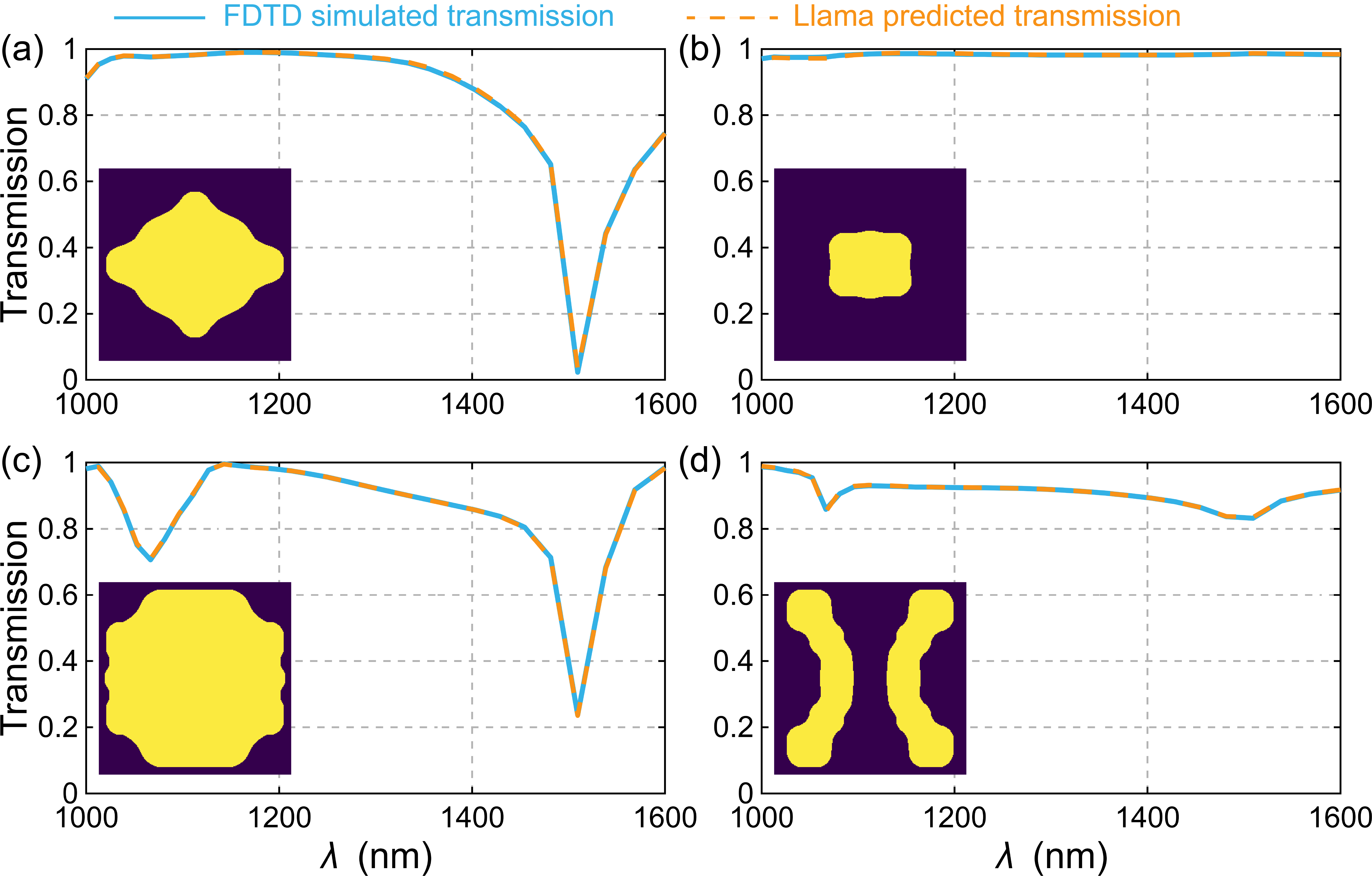} 
		\caption{Predicted and simulated transmission spectra for four grids from the test set. The corresponding control-point grids and MSE are: (a) [[0.411, 0.795, 0.126, 0.233], [0.876, 0.187, 0.209, 	0.911], [0.318, 0.479, 0.998, 0.826], [0.555, 0.820, 0.238, 0.058]], MSE = $7.8 \times 10^{-6}$. (b) [[0.156, 0.485, 0.350, 0.248], [0.391, 0.476, 0.083, 0.444], [0.041, 0.419, 0.524, 0.511], [0.695, 0.026, 0.690, 0.560]], MSE = $2.6 \times 10^{-6}$. (c) [[0.203, 0.155, 0.608, 0.655], [0.682, 0.541, 0.924, 0.898], [0.660, 0.610, 0.193, 0.065], [0.145, 0.508, 0.538, 0.098]], MSE = $3.6 \times 10^{-6}$. (d) [[0.049, 0.881, 0.405, 0.843], [0.288, 0.836, 0.375, 0.149], [0.736, 0.211, 0.728, 0.012], [0.471, 0.181, 0.914, 0.007]], MSE = $4.1 \times 10^{-6}$.}
		
	\end{figure}
	
	\subsection{Benchmarking}
	To assess how sensitive our workflow is to model choice, we fine-tuned eleven open-weight LLMs spanning three parameter bands “small” (< 7 B), “mid-size” (7-9 B), and “large” (> 9 B) on the same training-test split and fine-tuning setup, and summarized the resulting test-set MSEs in Figure. 3 b-d. Note that these regions are defined solely to show the feasibility of our method based on commonly used consumer-grade GPUs, rather than to align with definitions used in the machine learning community. Larger models are more sample-efficient during fine-tuning \citep{kaplan_scaling_2020}, and increasing epochs or fine-tuning data for larger models leads to diminishing returns \citep{zhang_when_2024}. Thus, the fine-tuning configuration used for mid-sized models is sufficient for other regions. In the mid-size model regime (Figure. 3b), accuracy generally improves with increasing size but not strictly monotonically: the 7B Qwen checkpoint reaches $4.0 \times 10^{-3}$, and the 9B Gemma variant levels off at $2.8 \times 10^{-3}$, indicating that entry-level GPUs can deliver spectra of acceptable fidelity. But Mistral 7B showed better MSE than 8B Llama variant, illustrating that architecture and internal design can outweigh simple parameter count increase. Accordingly, the size-accuracy gains discussed in Figure. 3b are best viewed as a trend rather than a strict rule. The small-model sweep (Figure. 3d) reinforces this point: Gemma-2-3B achieves $3.4 \times 10^{-3}$, whereas the tiny SmolLM2-0.1B variants drift above $14.7 \times 10^{-3}$. However, scaling further yields diminishing returns.  In particular, by enlarging Qwen-2.5 from 7B to 72B shaves only $1.2 \times 10^{-3}$ off the MSE yet stretches inference to almost 35 seconds and consumes the full 48 GB memory of a single NVIDIA L40S GPU (Figure. 3c). Gemma models rank first or second across all size bands, further suggesting that architectural priors outweigh raw parameter count in certain range. A plausible reason Gemma advances across models in our task is its larger, digit-friendly tokenizer, which represents decimals more regularly. Note that no deeper architectural tests are investigated here because the goal of our study is to provide a clear, out-of-the-box workflow that lets non-AI practitioners accelerate photonics design. Taken together, these bench-marks show that: (i) model selection can be guided by simple size thresholds rather than exhaustive hyper-parameter searches: changing the model size within the LLM family produces only modest accuracy shifts. In contrast, the DNN baseline shows a much larger spread across sizes. (ii) Gemma variants currently offer the best accuracy-to-cost ratio for rapid prototyping, and (iii) future gains are likely to come from designs that embed stronger numerical priors or VLMs rather than from continued parameter scaling alone.
	
	\begin{figure}
		\centering
		\includegraphics[width=\linewidth]{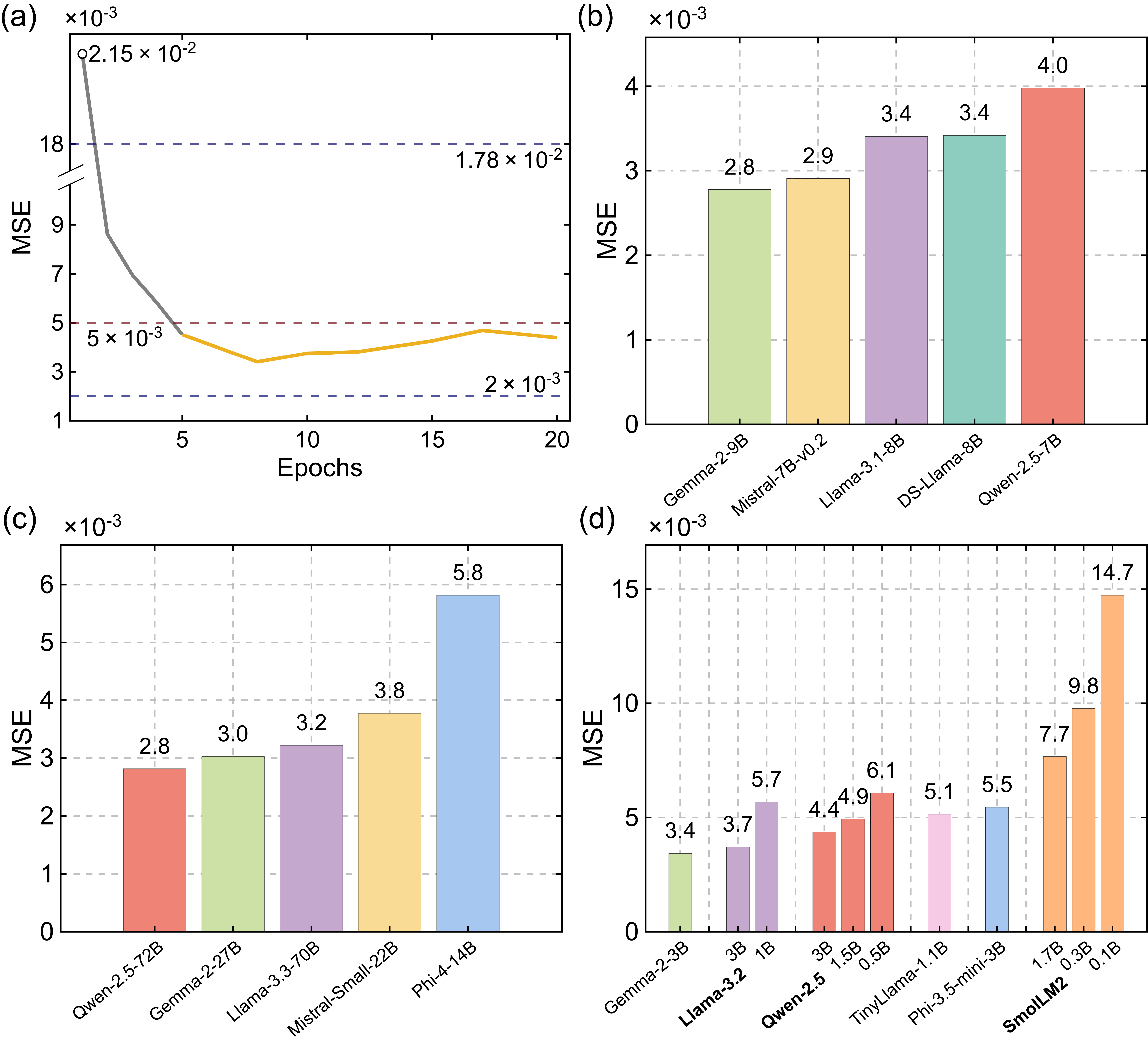} 
		\caption{(a) Test-set MSE for Llama-3.1-8B versus fine-tuning epochs. Although the MSE exceeds the $5 \times 10^{-3}$ tolerance line (red dashed line) during epochs 1-4 (grey curve), once fine-tuning reaches epoch 5 the orange curve remains consistently below this tolerance and only marginally above the $2.0 \times 10^{-3}$ benchmark reached by the best hand‑tuned eight‑layer DNNs (blue dashed line), indicating that predictive accuracy is largely insensitive to training length within a certain range. (b-d) MSE after eight-epoch LoRA fine-tuning for open-weight models grouped by size: (b) mid-size checkpoints (7-9B parameters). DS-Llama-8B stands for DeepSeek-distilled Llama-3.1-8B; (c) large models (> 9B); (d) small models (< 7B).}
		
	\end{figure}

	\subsection{Inverse design}
	Inverse metasurface design is fundamentally many-to-one: distinct geometries produce near-identical spectra, so a deterministic inverse network receives conflicting labels and its gradients cancel, stalling training, leading to non-convergence problems. Conventional remedies such as tandem networks, where an inverse generator is optimized through a frozen forward model \citep{liu_training_2018}, ease convergence but often collapse to a single prototype \citep{khaireh-walieh_newcomers_2023} and inherit the surrogate’s biases, thereby limiting design diversity \citep{chen_improved_2021}. Leveraging the intrinsic stochasticity of LLMs circumvents this problem. As sketched in Figure. 4a, we encode a 31-point transmission vector into the prompt “What's one grid of a metasurface that can produce the following spectrum: $[t_1, …, t_{31}]$”, invite the model to return “One possible grid would be $[[g_{11}, …, g{14}], …, [g{41}, …, g{44}]]$” and parse the tokens into the control-point grids. The deliberate phrasing “one possible” signals the multiplicity of valid answers explicitly, allowing the fine-tuned Llama-3.1-8B to learn that several candidates can generate similar spectra. After fine-tuning for 8 epochs, the Llama proposes a grid in about 0.9 seconds on a single RTX 2080 Ti GPU. Figure 4b (and more examples in Figure S1 of Supplementary Information) demonstrates four such inverse-designed meta-atoms: their FDTD-validated spectra (orange dashed) closely track the targets (blue solid) while their geometries differ markedly, confirming both fidelity and diversity without the need for techniques typically used in customized DNN approaches to mitigate non-convergence problems. Collectively, these results position LLMs as a fast, versatile alternative for inverse electromagnetic design. To compare to simple inverse baselines, we also implement a classical tandem inverse network, where an inverse network (spectrum to control points) is trained through a frozen forward network. Architecture details and representative results are provided in Figure S2 and S3 of Supplementary Information, with detailed observations from the comparison. 
	
	\begin{figure}
		\centering
		\includegraphics[width=\linewidth]{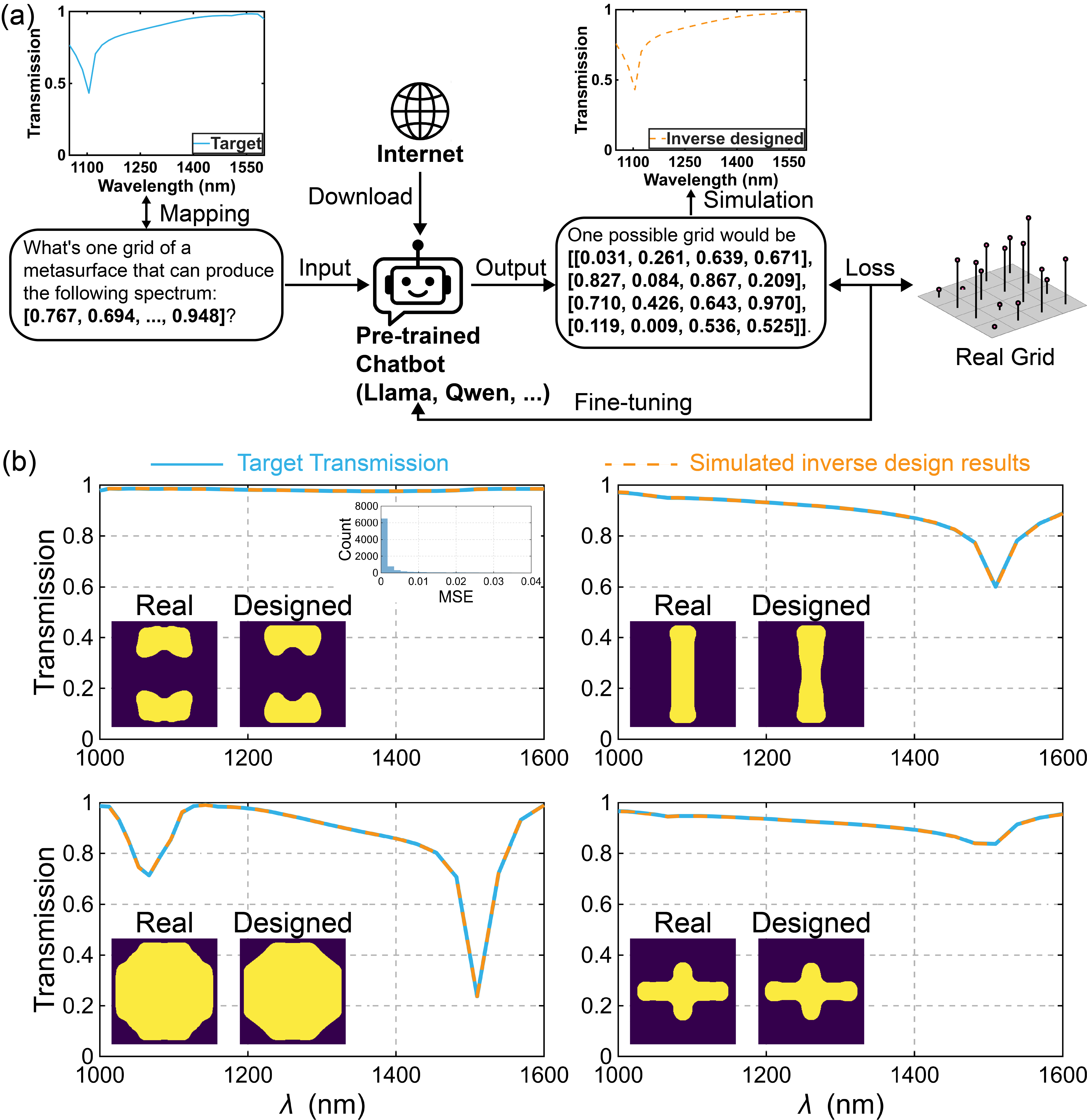} 
		\caption{(a) Workflow of the inverse-design stage. A target 31-point transmission spectrum is fed to the fine-tuned Llama-3.1-8B as a natural-language query of a corresponding grid; the model autoregressively returns a control-point grid that defines a candidate meta-atom. (b) Representative results for four unseen targets. The orange dashed lines are FDTD simulated results of inverse-designed metasurfaces. The corresponding inverse-designed grids and MSE are: top-left: [[0.550, 0.073, 0.906, 0.559], [0.324, 0.326, 0.831, 0.708], [0.916, 0.060, 0.517, 0.120], [0.023, 0, 0.249, 0.263]], MSE = $2.0 \times 10^{-7}$; top-right: [[0.360, 0.903, 0.903, 0.822], [0.419, 0.386, 0.377, 0.962], [0.744, 0.397, 0.391, 0.742], [0.890, 0.048, 0.259, 0.686]], MSE = $1.2 \times 10^{-6}$; bottom-left: [0.460, 0.289, 0.513, 0.473], [0.199, 0.641, 0.932, 0.866], [0.757, 0.956, 0.755, 0.282], [0.9120, 0.571, 0.547, 0.876]], MSE = $1.4 \times 10^{-6}$; bottom-right: [[0.964, 0.207, 0.656, 0.287], [0.777, 0.548, 0.192, 0.460], [0.181, 0.202, 0.218, 0.812], [0.303, 0.866, 0.496, 0.582]], MSE = $3.0 \times 10^{-7}$. The histogram within the top-left figure depicts the inverse-design test-set MSE distribution, showing that over 88\% of samples achieve an MSE below $1.0 \times 10^{-2}$.}
		
	\end{figure}

	\section{Conclusions}
	In summary, this work demonstrates that one-dimensional token-wise LLMs can serve as a practical “chat-to-chip” solution for both forward and inverse design of two-dimensional arbitrarily shaped metasurfaces without the need for vision models. Systematic benchmarking across widely used open-weight LLM checkpoints not only quantifies performance but also supplies a clear reference for future research. Collectively, these findings lower the barrier to entry for nanophotonic researchers who lack machine learning expertise and foreshadow a design paradigm in which LLMs drive rapid, automated exploration of increasingly complex metasurfaces and multifunctional electromagnetic devices.
	
	\paragraph{Research funding} This work was supported by the John L. and Genevieve H. McCain endowed chair professorship at the Pennsylvania State University.
	
	\paragraph{Author contribution} All authors have accepted responsibility for the entire content of this manuscript and consented to its submission to the journal, reviewed all the results and approved the final version of the manuscript. H. Z. and L. K. conceived the idea. H. Z. designed the experiments, developed the model code, performed the simulations. All authors contributed to the preparation of manuscript.
	
	\paragraph{Conflicts of interest} Authors state no conflict of interest.
	
	\paragraph{Data availability statement} The datasets generated and analysed during the current study are available from the corresponding author upon reasonable request. The source code for the fine-tuning workflow, data generation scripts, and analysis used in this study are available at \url{https://github.com/zhanghsh9/AbtryShapeTLLM}.
	
	
	\bibliographystyle{IEEEtranN}   
	\bibliography{references}       
	

\section*{Supplementary material}

# Supplementary Materials

## Chat to Chip: Large Language Model Based Design of Arbitrarily Shaped Metasurfaces

**Huanshu Zhang,[1] Lei Kang,[1] Sawyer D. Campbell,[1] and Douglas H. Werner[1, *]**

[1] *Department of Electrical Engineering, The Pennsylvania State University, University Park, PA 16802, USA.*
[*]Corresponding Author: dhw@psu.edu

**This SI includes:**

    **1. More examples of inverse design that show distinction between target shapes and designed shapes**
    **2. Tandem network results**
    **3. Observations between tandem networks and LLMs for inverse design**

## 1. More examples of inverse design

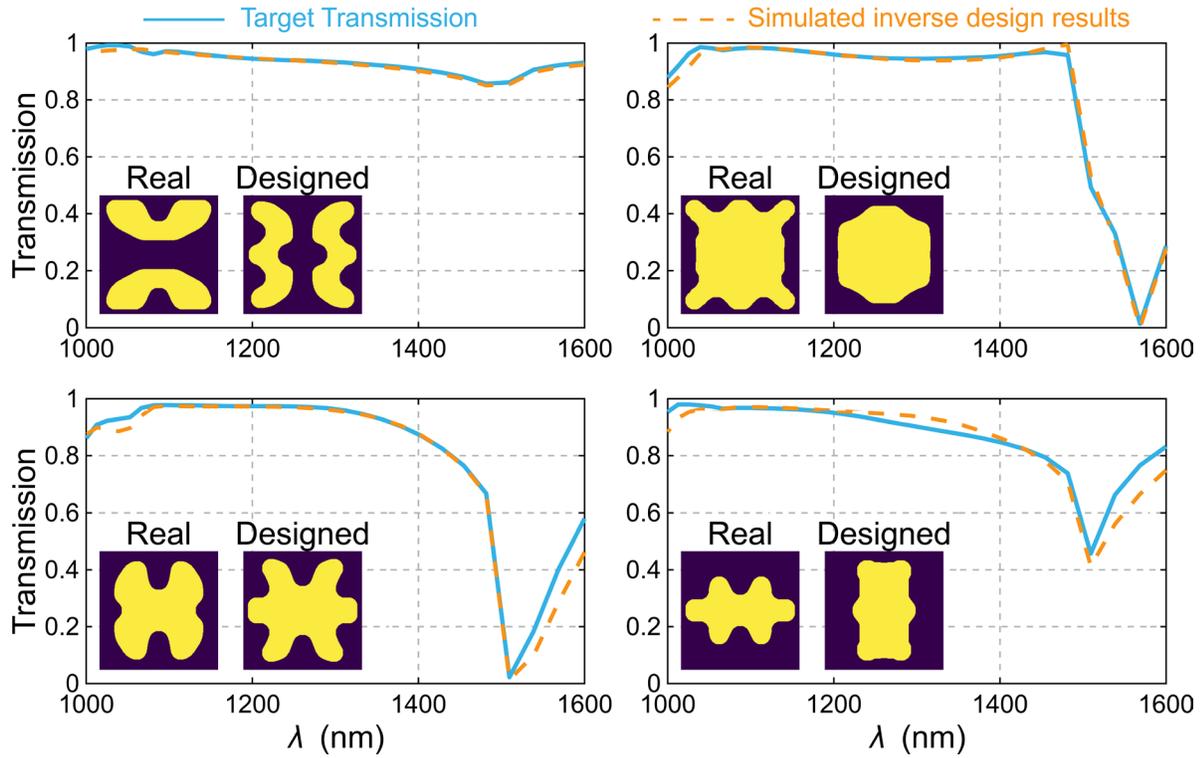

Figure S1. Representative results for four unseen targets. The orange dashed lines are FDTD simulated results of inverse-designed metasurfaces. The corresponding inverse-designed grids and MSE are: top-left: [[0.453, 0.285, 0.247, 0.896], [0.148, 0.960, 0.787, 0.045], [0.286, 0.335, 0.984, 0.063], [0.429, 0.859, 0.050, 0.453]], MSE = $6.6 \times 10^{-5}$; top-right: [[0.200, 0.006, 0.426, 0.290], [0.714, 0.453, 0.600, 0.606], [0.206, 0.632, 0.584, 0.563], [0.024, 0.622, 0.182, 0.765]], MSE = $3.1 \times 10^{-4}$; bottom-left: [[0.649, 0.506, 0.089, 0.160], [0.162, 0.589, 0.824, 0.082], [0.843, 0.396, 0.854, 0.593], [0.937, 0.792, 0.716, 0.887]], MSE = $1.3 \times 10^{-3}$; bottom-right: [[0.803, 0.217, 0.334, 0.287], [0.139, 0.278, 0.606, 0.454], [0.607, 0.347, 0.479, 0.406], [0.319, 0.422, 0.691, 0.693]], MSE = $1.5 \times 10^{-3}$. These results demonstrate that the LLM approach successfully mitigates the many-to-one non-convergence problem.

## 2. Tandem network results

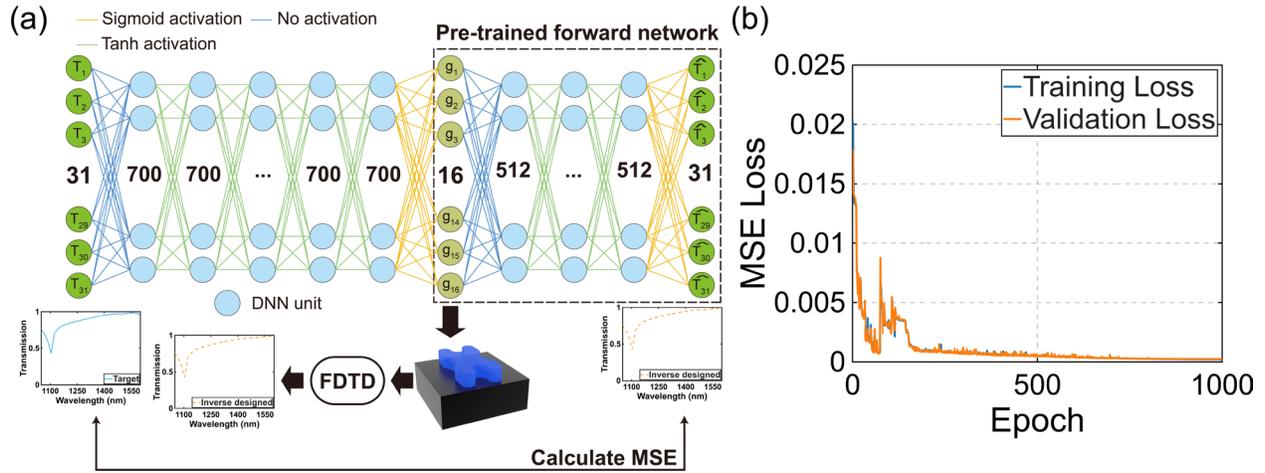

Figure S2. Tandem network structure and training loss. (a) A classical tandem network structure. Both forward and inverse networks are composed of multiple fully connected layers. The pre-trained forward network ingests the 4×4 control points flattened to a 16-dimensional vector, passes them through an initial fully connected expansion to 512 units, and then proceeds through a homogeneous stack of eight hidden layers, each of width 512, with a tanh applied after every hidden affine transformation; a final linear layer followed by a component-wise sigmoid yields a 31-dimensional spectral output constrained to [0, 1]. The MSE of this forward network on the test set is $2.0 \times 10^{-3}$. Conversely, the inverse network accepts a 31-element spectrum, expands to 700 units, and traverses six hidden layers of width 700 with the same tanh activation after each hidden layer, concluding with a linear projection and component-wise sigmoid that returns a 16-dimensional vector corresponding to the flattened 4×4 control points, also bounded to [0, 1]. The loss function calculates the MSE between the target transmission and the predicted transmission and only uses this loss to update the weights of the inverse network while keeping the forward network unchanged. The final MSE′, which is defined as the MSE between the target transmissions and the predicted transmissions by the forward network, is $2.4 \times 10^{-4}$. This figure is reproduced with permission from [Zhang, Huanshu, *et al.* "Fixed-attention mechanism for deep-learning-assisted design of high-degree-of-freedom 3D metamaterials," *Optics Express* 33.9 (2025): 18928-18937]. Copyright 2025 Optical Society of America. (b) Learning curve for the backward training process.

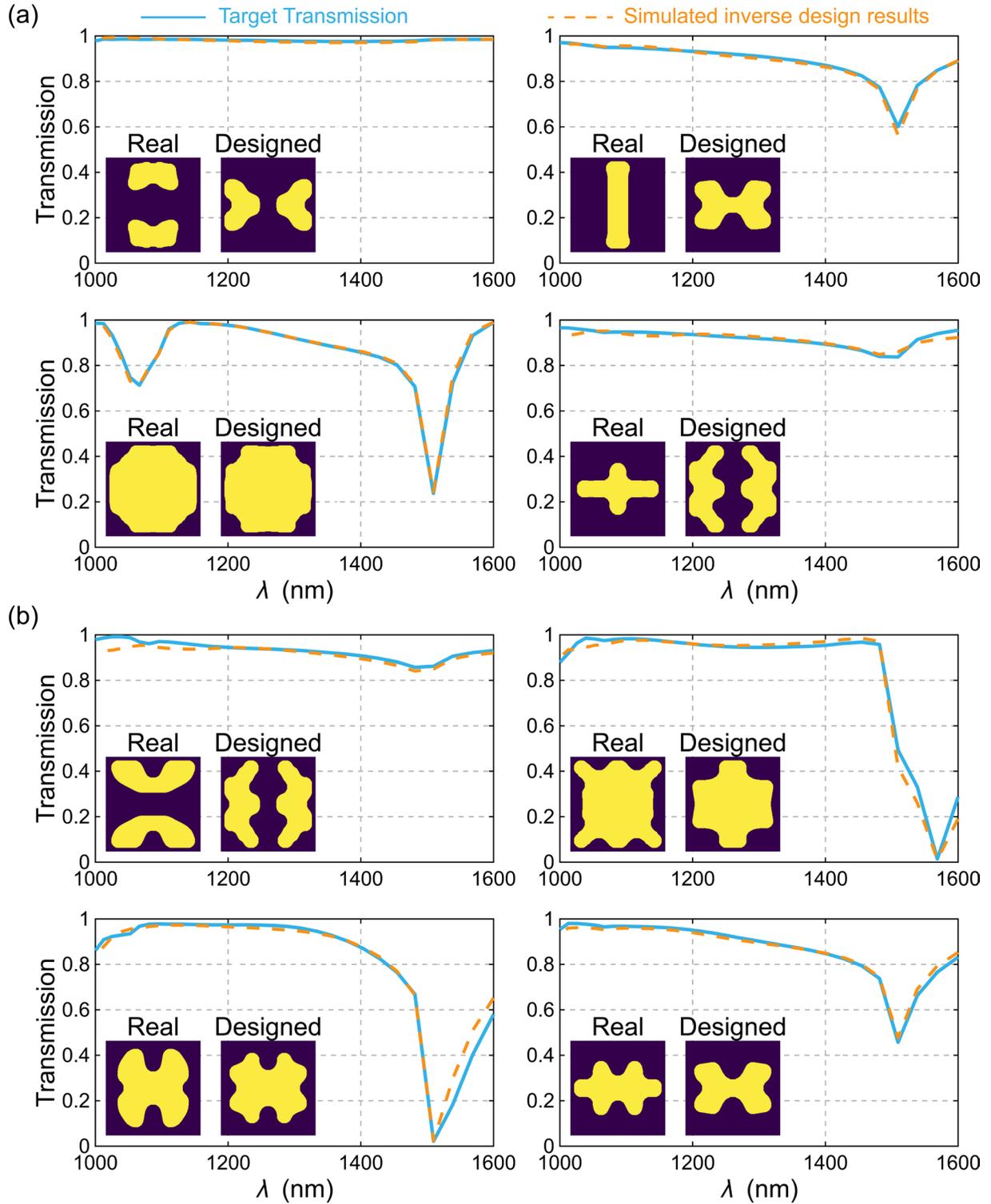

Figure S3. Representative results for some unseen targets. The orange dashed lines are FDTD simulated results of inverse-designed metasurfaces. (a) The corresponding inverse-designed grids and MSE are: top-left: [[1.000, 1.000, 1.000, 1.000], [1.000, 0.198, 0.305, 0.037], [0.622, 1.000,

0.284, 0.559], [0.222, 0.732, 1.000, 0.005]], MSE = $1.9 \times 10^{-5}$, the MSE of the same design from LLM is $2.0 \times 10^{-7}$; top-right: [[1.000, 1.000, 1.000, 1.000], [1.000, 0.127, 0.315, 0.008], [0.002, 1.000, 0.775, 0.0525], [0.268, 0.497, 1.000, 0.569]], MSE = $1.0 \times 10^{-4}$, the MSE of the same design from LLM is $1.2 \times 10^{-6}$; bottom-left: [[1.000, 1.000, 1.000, 1.000], [1.000, 0.391, 0.973, 0.024], [0.304, 1.000, 0.053, 0.588], [0.750, 0.931, 1.000, 0.758]], MSE = $7.4 \times 10^{-5}$, the MSE of the same design from LLM is $1.4 \times 10^{-6}$; bottom-right: [[1.000, 1.000, 1.000, 1.000], [1.000, 0.408, 0.554, 0.058], [0.699, 1.000, 0.218, 0.643], [0.281, 0.886, 1.000, 0.028]], MSE = $2.2 \times 10^{-4}$, the MSE of the same design from LLM is $3.0 \times 10^{-7}$. (b) The corresponding inverse-designed grids and MSE are: top-left: [[1.000, 1.000, 1.000, 1.000], [1.000, 0.376, 0.524, 0.048], [0.627, 1.000, 0.280, 0.604], [0.304, 0.875, 1.000, 0.030]], MSE = $6.0 \times 10^{-4}$, the MSE of the same design from LLM is $6.6 \times 10^{-5}$; top-right: [[1.000, 1.000, 1.000, 1.000], [1.000, 0.256, 0.520, 0.726], [0.177, 1.000, 0.983, 0.995], [0.108, 0.986, 1.000, 0.346]], MSE = $8.5 \times 10^{-4}$, the MSE of the same design from LLM is $3.1 \times 10^{-4}$; bottom-left: [[1.000, 1.000, 1.000, 1.000], [1.000, 0.237, 0.761, 0.074], [0.001, 1.000, 0.8710, 0.592], [0.017, 0.593, 1.000, 0.553]], MSE = $1.1 \times 10^{-3}$, the MSE of the same design from LLM is $1.3 \times 10^{-3}$; bottom-right: [[1.000, 1.000, 1.000, 1.000], [1.000, 0.224, 0.387, 0.011], [0.001, 1.000, 0.803, 0.144], [0.104, 0.503, 1.000, 0.835]], MSE = $1.8 \times 10^{-4}$, the MSE of the same design from LLM is $1.5 \times 10^{-3}$. These results demonstrate that the classical tandem network approach successfully mitigates the many-to-one non-convergence problem.

## 3. Observations between tandem networks and LLMs for inverse design

The LLM-based inverse designer exhibits greater solution diversity than the tandem baseline while lacking fidelity in certain structures. In the tandem setting, predicted control points frequently saturate at the upper bound (repeating 1.000 entries across large subblocks), which is characteristic of boundary clamping and partial mode collapse when an inverse network is trained solely through a frozen forward network under a bounded output layer. By contrast, the LLM produces interior-valued, heterogeneous grids that avoid clamping. The LLM's spread of solutions is consistent with the intrinsically many-to-one nature of inverse metasurface design. Taken together, these observations indicate that the LLM preserves geometric degrees of freedom and achieves a favorable fidelity-diversity-AI-knowledge trade-off without collapsing to boundary solutions, qualities that are desirable for practical inverse design workflows.


\end{document}